\documentclass{aa}
\usepackage[dvips]{graphics}
\usepackage{latexsym}
\def\gtsima{$\; \buildrel > \over \sim \;$}
\def\ltsima{$\; \buildrel < \over \sim \;$}
\def\prosima{$\; \buildrel \propto \over \sim \;$}
\def\gsim{\lower.5ex\hbox{\gtsima}}
\def\lsim{\lower.5ex\hbox{\ltsima}}
\def\simgt{\lower.5ex\hbox{\gtsima}}
\def\simlt{\lower.5ex\hbox{\ltsima}}
\def\simpr{\lower.5ex\hbox{\prosima}}

\def\h1{$h^{-1}$}
\def\beq{\begin{equation}}
\def\eeq{\end{equation}}
\begin{document}
\title{
The manifold spectra and morphologies of EROs\thanks{Based 
on observations made at the European Southern Observatory,
Paranal, Chile (Program 70.A-0140(B)).}}
\author{
	A. Cimatti \inst{1}
	\and
	E. Daddi \inst{2}
	\and
	P. Cassata \inst{3}
	\and 
	E. Pignatelli \inst{4}
	\and
	G. Fasano \inst{4}
	\and
	J. Vernet \inst{1}
	\and 
	E. Fomalont \inst{5}
	\and 
	K. Kellermann \inst{5}
	\and
	G. Zamorani \inst{6}
	\and
	M. Mignoli \inst{6}
	\and
	L. Pozzetti \inst{6}
	\and
	A. Renzini \inst{2} 
	\and
	S. di Serego Alighieri \inst{1}
	\and 
	A. Franceschini \inst{3}
	\and 
	E. Giallongo \inst{7}
	\and
	A. Fontana \inst{7}
}
\institute{ 
INAF - Osservatorio Astrofisico di Arcetri, Largo E. Fermi 5, I-50125 
Firenze, Italy 
\and European Southern Observatory, Karl-Schwarzschild-Str. 2, D-85748, 
Garching, Germany
\and Dipartimento di Astronomia, Universit\`a di Padova, Vicolo
dell'Osservatorio, 2, I-35122 Padova, Italy 
\and INAF - Osservatorio Astronomico di Padova, Vicolo dell'Osservatorio, 5, 
I-35122 Padova, Italy
\and National Radio Astronomy Observatory, Charlottesville, VA 22903, USA
\and INAF - Osservatorio Astronomico di Bologna, via Ranzani 1, I-40127, 
Bologna, Italy
\and INAF - Osservatorio Astronomico di Roma, via dell'Osservatorio 2, 
Monteporzio, Italy
}
\offprints{Andrea Cimatti, \email{cimatti@arcetri.astro.it}}
\date{Received ; accepted }
\abstract{
Deep VLT optical spectroscopy, HST+ACS ({\sl GOODS}) imaging and VLA 
observations are used to unveil the nature of a complete sample of 
47 EROs with $R-K_{\rm s}>5$ and $K_{\rm s}<20$. The spectroscopic redshift 
completeness is 62\%. Morphological classification was derived for each 
ERO through visual inspection and surface brightness profile fitting. 
Three main ERO morphological types are found: E/S0 galaxies ($\sim 30-37$\%), 
spiral-like ($\sim 24-46$\%) and irregular systems ($\sim 17-39$\%). 
The only ERO detected in the radio is likely to host an obscured AGN. The 
average radio luminosity of the star-forming EROs undetected in the radio 
implies star formation rates of the order of $\sim$33 M$_{\odot}$yr$^{-1}$.
The colors, redshifts and masses of the E/S0 galaxy subsample imply 
a minimum formation redshift $z_{\rm f}\sim 2$. With this $z_{\rm f}$
there is enough time to have old and massive 
stellar spheroids already assembled at $z\sim 1$. We verify that the 
$R-K_{\rm s}$ vs. $J-K_{\rm s}$ color diagram is efficient in segregating old 
and dusty-star-forming EROs. 
} 
\titlerunning{Manifolds of EROs}
\authorrunning{Cimatti et al.}  \maketitle

\section{Introduction}

Extremely Red Objects (EROs) are important probes of galaxy evolution.
On one hand, they allow to constrain the formation epoch, the evolutionary 
pattern and the clustering of the oldest early-type galaxies at $z$ \gtsima 1 
(e.g. Daddi et al. 2000a,b; McCarthy et al. 2001; Cimatti et al. 
2002a; Smith et al. 2002; Firth et al. 2002; Moustakas \& Somerville 
2002; Roche et al. 2002, 2003; Saracco et al. 2003). On the other 
hand, they allow to select 
dusty star-forming galaxies and obscured AGN in a way complementary to 
submillimeter surveys (Cimatti et al. 1998; Afonso et al. 2001;
Mohan et al. 2002; Cimatti 
et al. 2002a; Franceschini et al. 2002; Brusa et al. 
2002). Although recent VLT spectroscopy shed new light on moderately
bright EROs with $K_{\rm s}$\ltsima 19 (Cimatti et al. 2002a), the 
characteristics of fainter EROs is still unclear mainly because of their
weakness which makes spectroscopy very challenging even with 8-10m-class 
telescopes. However, as different morphologies are expected in case of 
spheroids or star-forming systems, HST imaging can be used to complement 
spectroscopy for the faint ERO population (e.g. Moriondo et al. 2000; 
Yan \& Thompson 2003). 
The main open questions about EROs include: the relative
fraction of different ERO types, the nature and star formation rates
of dusty EROs, the number density of the oldest high-$z$
spheroids selected as EROs, the fraction of obscured AGN hosted by
EROs. In this paper we present the main results of 
very deep VLT spectroscopy, HST imaging and VLA observations aimed at 
investigating the nature of a complete sample of 47 EROs with $K_{\rm s}<20$. 
The properties of the sample will be presented in a 
companion paper (Cassata et al. 2003, in preparation). $H_0=70$ km s$^{-1}$ 
Mpc$^{-1}$, $\Omega_{\rm m}=0.3$ and $\Omega_{\Lambda}=0.7$ are adopted.

\section{The sample and the observations} 

The complete sample (already described by Cimatti et al. 2002a) is made by 
the 47 EROs with $R-K_{\rm s}>5$ and $K_{\rm s}<20$ present in the 32.2 arcmin$^2$ 
area of the {\it Chandra Deep Field South} (CDFS; Giacconi et al. 2001) 
covered by the K20 survey (Cimatti et al. 2002b). One object
was removed from the original sample because it was found to be a blend 
of two objects both with $K_{\rm s}>20$.

New VLT-UT4+FORS2 multi-object optical spectroscopy was made in November 
2002 with the red-sensitive CCD array. The total integration time 
was 7.8 hours during 0.7$^{\prime\prime}$-1.0$^{\prime\prime}$ seeing
conditions, with 1.0$^{\prime\prime}$ wide slits, grism 300I (R$\sim$660) 
and dithering of the targets along the slits. The mask included 13 EROs 
of the K20/CDFS sample that were previously unidentified or not observed.
The spectra were reduced and analyzed with the methods described
Cimatti et al. (2002a, 2002b). VLT spectroscopy was complemented with 
the HST+ACS imaging in the $BViz$ bands taken in the {\sl GOODS/HST 
Treasury Program} (Giavalisco et al. 2003). 

Deep radio data were obtained with the VLA at 1.4 GHz in 1999 October
with a bandwidth of 25 MHz in each of two circular polarizations, and 
at 4.8 GHz in 1999 Feb with a bandwidth of 100 MHz in each of two 
circular polarizations,
for 12-hours total at both frequencies (Kellermann et al. 2003, in
preparation). The rms sensitivities were 8.6 $\mu$Jy and 9.6 $\mu$Jy 
at 1.4 and 4.8 GHz, respectively, each with a
full-width half-power resolution of $3.5^{\prime\prime}$. The reductions were
made using AIPS and each field was self-calibrated. The position and 
flux density of each source with a peak flux density greater than 
3.5-rms was determined from a Gaussian model fit of the source.  

\section{VLT spectroscopy and HST imaging results}

The 13 ERO spectra provided 9 new redshift measurements:
7 star-forming galaxies at $z\sim 1.2-1.6$ with [OII]$\lambda$3727 
emission, and 2 early-type objects at $z\sim 1.5$. When added to the 
previous identifications of Cimatti et al. (2002a), the spectroscopic 
redshift completeness for the EROs in the K20/CDFS area increases to 
81\% (21/26) ($K_{\rm s}<19.2$) and 62\% (29/47) ($K_{\rm s}<20.0$). 
Photometric redshifts for the 18 spectroscopically unidentified EROs were 
derived using the ESO/{\sl GOODS} VLT+FORS1 $BVRIz$ and the new
VLT+ISAAC $JHK_{\rm s}$ public imaging and the $Hyperz$ software (Bolzonella et 
al. 2000). The agreement between spectroscopic and photometric redshifts
is good and does not depend on the ERO class, with $\Delta z = 
\langle z_{\rm spec}-z_{\rm phot}\rangle$=0.001 and $\sigma(\Delta z)$=0.16 (29
EROs). Figure 1 shows the redshift distributions divided into morphological
classes (see Section 4). 
\begin{figure}[ht]
\resizebox{\hsize}{!}{\includegraphics{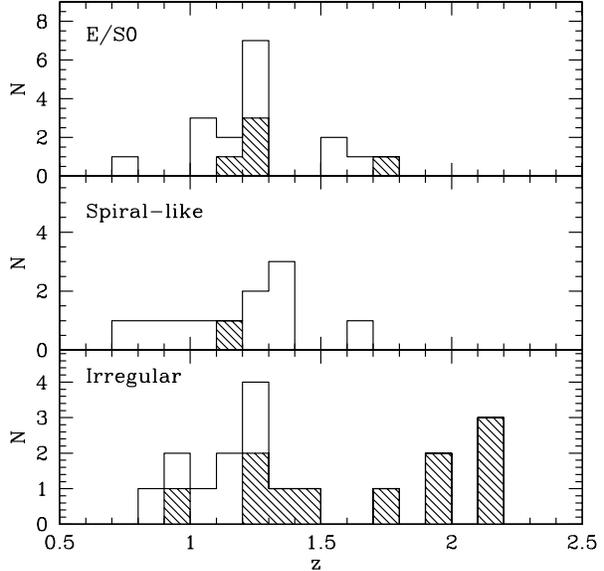}}
\caption{\footnotesize 
The redshift distributions. The shaded histograms indicate the
photometric redshifts. 
}
\label{fig:plot}
\end{figure}

The morphological type of EROs was first estimated by visual
inspection of the ACS images and isophotal maps in the four
available bands ($BViz$). One ERO (not observed spectroscopically) 
is unresolved and has colors consistent with being a M star. 
Since in some cases the spatial resolution and/or the S/N ratio 
turned out to be limited, we just assigned each object to one of 
the following broad classes: E/S0 galaxies (class 1), spiral-like 
objects (class 2) and irregular/interacting systems (class 3). 
Ambiguous cases were flagged by introducing the intermediate classes 
1.5 and 2.5. Thus, we used 
the GASPHOT software (Pignatelli and Fasano 1999) to carry out 
a quantitative analysis on the $z$-band ACS images (filter F850LP) by 
fitting the surface brightness profile of each ERO with the Sersic 
law ($SB\propto r^{1/n}$), convolved with the average PSF extracted 
from the stars in the field. Being the bulk of the sample at $z\sim 1$, 
the filter F850LP is close to the rest-frame $B$-band, thus minimizing 
the effects of dust extinction and morphological K-correction for the 
available set of ACS filters. EROs belonging to the class 1.5 and 
having $n\geq$2.5 or $\leq$1.5 were assigned to the class 1 or 2, 
respectively, while the cases with 1.5$< n <$2.5 were assigned to 
intermediate class 1.5. Instead, the Sersic 
index showed to be ineffective in disentangling classes 2 and 3, 
making advisable to leave unchanged most of the dubious cases 
with flag 2.5 (10 objects). Indeed, for six of these objects 
the fitting algorithm failed to converge, due to the extremely 
irregular and noisy image structure. These objects are not reported 
in the lower panel of Fig.2, where the final classification is 
plotted versus the Sersic index. We note that the larger number of 
class-2.5 vs. class-1.5 objects and the higher redshifts 
of ``irregular'' EROs (Fig.1) may be due to morphological K-correction 
effects (e.g. Marcum et al. 2001; Windhorst et al. 2002).
By taking into account the previous uncertainties, the ranges of 
morphological fractions are 30--37\% (class 1 -- class 1+1.5), 24--46\% 
(class 2 -- class 2+2.5) and 17--39\% (class 3 -- class 2.5+3). The
fractions are in reasonable agreement with the recent results by 
Moustakas et al. (2003) and Gilbank et al. (2003), and do not 
significantly change as a function of the limiting $K_{\rm s}$ magnitude. 
Fig.2 (top panel) and Fig.3 show that a rather good agreement is present
between the morphological and spectroscopic classes. 
\begin{figure}[ht]
\resizebox{\hsize}{!}{\includegraphics{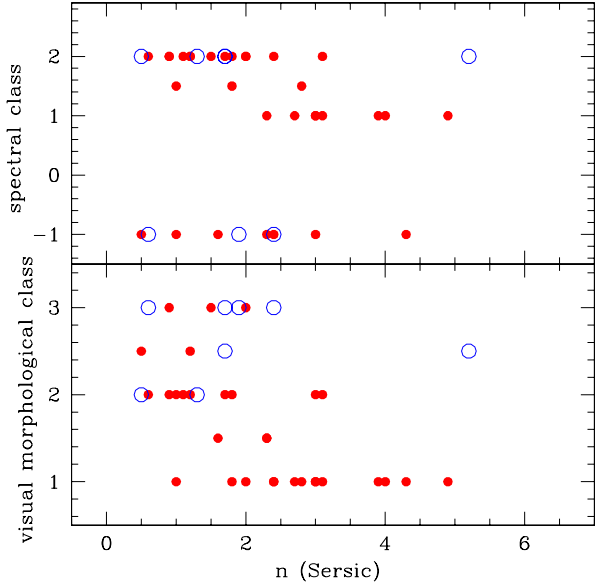}}
\caption{\footnotesize 
The Sersic index $n$ vs. the spectral (top) and visual
morphological class (bottom). Filled symbols: reliable fit 
of $n$; open symbols: uncertain fit of $n$. Spectral classes: -1 = 
unidentified or unobserved, 1 = passive early-type, 1.5 = 
early-type + weak [OII]$\lambda$3727 emission, 2 = star-forming. 
Visual morphological classes: 1 = E/S0, 2 = spiral-like, 3 = irregular.
}
\end{figure}

\begin{figure}[ht]
\resizebox{\hsize}{!}{\includegraphics{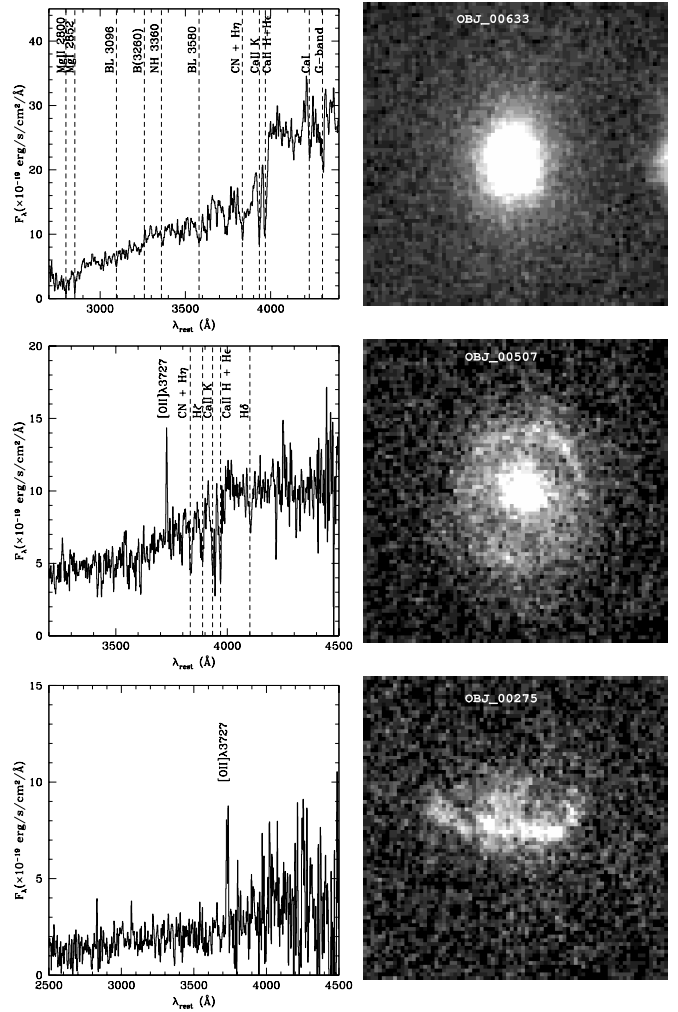}}
\caption{\footnotesize 
Examples of EROs VLT spectra and HST morphologies (ACS filter F850LP, 
box size = $4\times4$ arcsec). From top: E/S0 ($z=1.096$), spiral-like 
($z=1.222$), irregular ($z=1.221$).
}
\label{fig:plot}
\end{figure}
\section{Dusty star-forming EROs and obscured AGN}

Only one object was detected at $>5\sigma$ level ($S>50$ $\mu$Jy)
in the 1.4 GHz radio source catalog of the CDFS (Kellermann et al. 
2003, in preparation). Our radio detection rate (1$^{+2.30}_{-0.83}$/46 
at $1\sigma$ poissonian confidence level) is compatible with the 
preliminary results of the {\it Phoenix Deep Survey} (Hopkins et al. 
2003) where $\sim$6\% of a sample of $\sim$425 EROs ($K_{\rm s}<20$, 
$R-K_{\rm s}>5$) are detected with fluxes $>60$ $\mu$Jy ($5\sigma$) 
(J. Afonso, private communication), with the results 
of Smail et al. (2002) (3/50 EROs with $K_{\rm s}<20$, $R-K_{\rm s}>5.3$ and
$S>50$ $\mu$Jy at 1.4 GHz) and Roche et al. (2002) (1/31 EROs with
$K_{\rm s}<19.5$, $R-K_{\rm s}>5$ and $S>50$ $\mu$Jy at 1.4 GHz).

The ERO detected in the radio is an emission line object with 
[OII]$\lambda$3727 at $z=1.610$
and a spiral-like morphology. The flux densities at 1.4 GHz and 
4.8 GHz are 106.8$\pm$8.6 $\mu$Jy and 155.1$\pm$9.6 $\mu$Jy respectively, 
implying a flat/inverted spectrum with $\alpha_{\rm radio}=+0.3$ ($S(\nu)\propto 
\nu^{\alpha}$). The object is also present in the X-ray catalog of
Alexander et al. (2003) (RA=03:32:10.80, Dec= -27:46:27.6, J2000) with fluxes 
F(0.5-2 keV)=8.67$\times 10^{-17}$ cgs and F(0.5-8 keV)=3.08$\times 
10^{-16}$ cgs, corresponding to luminosities of 
1.4$\times 10^{42}$ erg s$^{-1}$ and 5$\times 10^{42}$ erg s$^{-1}$ 
respectively. Its X-ray luminosity, $\alpha_{\rm radio}$ and location in typical
segregation diagrams (e.g. $\alpha_{\rm radio}$ vs. radio flux, optical 
flux vs. 0.5-8 keV flux; Bauer et al. 2002) are more suggestive of 
a Type 2 AGN than a starburst galaxy. The lack of broad MgII$\lambda$2800 
emission implies obscuration of the active nucleus. From the X-ray hardness 
ratio we derive log(N$_{\rm H}$)$<$ 22.7 cm$^{-2}$ (assuming a photon index 
$\Gamma=1.7$). 

Although EROs are often found as faint counterparts of X-ray selected 
sources (e.g. Yan et al. 2003), the fraction of $K$-selected EROs 
with X-ray emission is rather small (e.g. Alexander et al. 2002; 
14$^{+11}_{-7}$\% at $K<$20.1; Roche et al. 2003; $\sim$8\% at 
$K_{\rm s}<20$). Taking into account the only other X-ray source 
detected in our ERO sample (Brusa et al. 2002), the fraction of X-ray 
emitting EROs amounts to 2/46 (1.5-10\% at 1$\sigma$ poissonian 
confidence level), consistently with the above mentioned results. 

Although only one ERO has been detected significantly in the radio, 
a mean flux density of the undetected sources was 
estimated by comparing the flux density distribution of the
non-detected ERO sources with that expected from noise and confusion.
The radio intensity distribution at 1.4 GHz for the 28 non-detected
spiral-like + irregular ERO sources has a rms of $9.3~\mu$Jy and an 
average value of $3.5~\mu$Jy compared with an rms of $8.9~\mu$Jy and 
average value of $-0.8~\mu$Jy for the control sample of 270 random 
locations in the field.  From this excess we infer that the average flux 
density of ERO sources at 1.4 GHz is $3.5\pm 1.7~\mu$Jy. At 5 GHz we estimate 
an average flux density of $<4~\mu$Jy. For the 17 E/S0 galaxies we 
obtained limit of $<3~\mu$Jy at both 1.4 and 5 GHz.

Adopting a radio spectral index $\alpha=-0.6$, the average rest-frame 1.4 GHz 
luminosity is $2.7\times10^{22}$ W Hz$^{-1}$ for the average redshift
of the 28 undetected star-forming EROs ($\langle z\rangle=1.337$). 
Following Condon (1992)
and Haarsma et al. (2000) (with Salpeter IMF and 0.1-100 M$_{\odot}$ 
mass range), we derive SFR$\sim$ 33 M$_{\odot}$ yr$^{-1}$. The estimated 
SFRs are consistent with those obtained by Brusa et al. (2002) using 
the X-ray emission (5--44 M$_{\odot}$ yr$^{-1}$). Following Kennicutt 
(1998), the expected average far-IR luminosity is $\sim 1-2\times10^{11}$ 
L$_{\odot}$ corresponding to 850$\mu$m observed fluxes below 1 mJy, thus 
explaining the low detection rate of $K_{\rm s}<20$ EROs with submm/mm 
continuum observations (e.g. Mohan et al. 2002). In comparison, the 
[OII]$\lambda$3727 luminosity of all the spectroscopically identified 
star-forming EROs in the CDFS provides an average SFR=2.8 M$_{\odot}$ 
yr$^{-1}$. If this difference with the radio is solely due to dust 
extinction, the required average reddening is $E(B-V)\sim0.5$. 
This amount of reddening is significantly larger than in Lyman-break
galaxies (median $E(B-V)\sim0.17$, range $E(B-V)\sim0.1-0.3$; 
Giavalisco 2002 and references therein).
The average SFR(radio) implies a cosmic SFR density of EROs at $z\sim1.3$ 
of $>$0.008 M$_{\odot}$ yr$^{-1}$ Mpc$^{-3}$ (adopting the comoving volume 
between $z_{\rm min}=0.726$ and $z_{\rm max}=2.185$), corresponding to about $>$12\% 
of the total budget at that redshift (e.g. Somerville et al. 2001). This is 
strictly a lower limit because our sample is limited to $K_{\rm s}<20$ and
due to the conservative volume adopted between $z_{\rm min}$ and $z_{\rm
max}$. 

\section{The oldest and massive spheroids at $z$ \gtsima 1}

The reddest spheroidals selected as EROs represent the oldest envelope 
of the E/S0 population at $z$ \gtsima 1 and are crucial to constrain 
the earliest epoch of massive system formation.
\begin{figure}[ht]
\resizebox{\hsize}{!}{\includegraphics{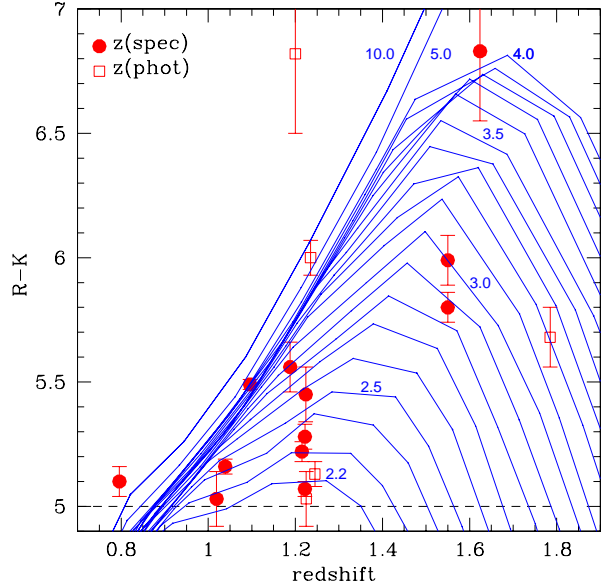}}
\caption{\footnotesize 
$R-K_{\rm s}$ vs. $z$ for the EROs with E/S0 morphology. Continuous lines 
indicate the predicted colors for different formation redshifts 
(indicated by the numbers) and $\tau=0.3$ Gyr.
}
\label{fig:plot}
\end{figure}
Taking into account the non-poissonian fluctuations due to clustering 
(Daddi et al. 2000a; Roche et al. 2002), the observed surface density of 
E/S0 galaxies (morphological class 1-1.5) is $0.50^{+0.37}_{-0.16}$ 
arcmin$^{-2}$ ($1\sigma$) ($0.50^{+1.50}_{-0.26}$ at $2\sigma$). 
The Pure Luminosity Evolution (PLE) model of Daddi et al. (2000b) with 
a star formation $e$-folding time $\tau=0.3$ Gyr and a formation 
redshift $z_{\rm f}\sim2.5$ predicts surface densities of $\sim$2 
arcmin$^{-2}$ and $\sim$0.9 arcmin$^{-2}$ for the 2MASS local early-type 
(E+S0+ some contamination by Sa galaxies; Kochanek et al. 2001) and the local 
elliptical (E only; Marzke et al. 1994) luminosity functions 
respectively. Although the large uncertainties on the surface density 
prevent a stringent comparison with the PLE model, we note that the 
observed density of E/S0 galaxies at 0.8\ltsima$z$\ltsima1.6 is 
still formally consistent with the above PLE predictions within 
1-2$\sigma$. The question of high-$z$ early-type galaxies is still
controversial (e.g. Treu 2003 and references therein), and large 
area surveys are required to overcome the effects of the strong clustering. 
Figure 4 shows a significant spread in the formation 
redshifts of the identified E/S0 galaxies (2.2\ltsima$ z_{\rm f}$\ltsima4.0 
for $\tau=0.3$ Gyr). If $\tau=0.1$ Gyr, the range is 2\ltsima$z_{\rm f}$\ltsima3.
From the lowest acceptable formation redshift ($z_{\rm f} \sim 2.0$) to 
$z\sim1$ there is enough time to reach an age of $\sim$2.5 Gyr, compatible 
with that estimated from the average spectrum and from the 4000~\AA~ 
continuum break amplitude of old EROs at $z\sim1$ (Cimatti et al. 2002a),
and within the range of the ages of $z>1$ spheroidal galaxies ($\sim$1.5-
3.5 Gyr) estimated with different spectral synthesis models (e.g. Nolan 
et al. 2001 and references therein). The rest-frame near-IR luminosities
of old EROs in our sample are in the range of -24.5 and -27 (Pozzetti
et al. 2003), and their stellar masses are $\sim 10^{11}$--$10^{12}$ 
M$_{\odot}$ (Fontana et al. 2003, in preparation).

The morphology, spectral properties, luminosities, inferred ages, 
formation redshifts
and stellar masses imply the existence of a substantial population of
old, passively evolving and fully assembled massive spheroids at
$z\sim1-1.5$ which requires that major episodes of massive galaxy formation 
occurred at $z$\gtsima2. Daddi et al. (2003) uncovered a population of 
massive and clustered star-forming galaxies at $z\sim2$ which may 
represent the long sought-for progenitors of massive spheroids.

\section{Testing the photometric segregation of EROs}

We used our sample to test the segregation of morphologically classified
EROs in the $R-K_{\rm s}$ -- $J-K_{\rm s}$ diagram proposed by Pozzetti \& Mannucci
(2000) (Fig. 5). 
The colors were derived from the new public ESO {\sl GOODS} imaging 
taken with the VLT FORS1 ($R$) and ISAAC ($JK_{\rm s}$).
\begin{figure}[ht]
\resizebox{\hsize}{!}{\includegraphics{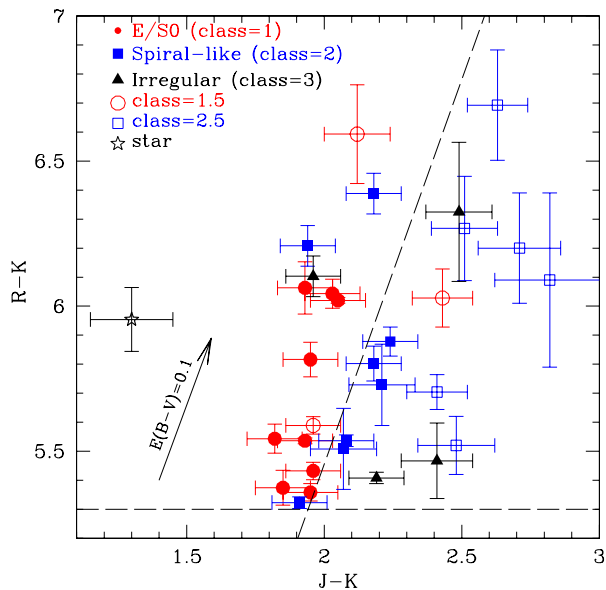}}
\caption{\footnotesize 
The segregation color - color diagram. The E(B-V) vector is 
relative to $z=1.1$ and the Calzetti extinction curve.
}
\end{figure}
The relation used in Fig. 5 (dashed line)
to define the boundary line between ``old'' 
and ``dusty'' EROs is $J-K_{\rm s}$=0.34($R-K_{\rm s}$)+0.09. This relation was 
derived specifically for the set of $RJK_{\rm s}$ filters of our photometry,
and is valid for $R-K_{\rm s}>$5.3 (Pozzetti \& Mannucci 2000). 

Two main results are evident from Fig.5: {\sl (1)} the E/S0 galaxies 
with secure morphological classification (class=1) and most of star-forming 
EROs lie in their expected regions, {\sl (2)} a few star-forming EROs 
(classes=2-3), however, are found in the region of E/S0 galaxies. 
If we consider only the EROs with reliable morphological classification, 
the contamination of the E/S0 region is 4/15 (27\%), whereas 4/18 (22\%) 
star-forming EROs do not fall in their expected locus. The contamination 
of the E/S0 region may be due to star-forming galaxies with a substantial 
underlying old stellar population which dominates the colors.
Our results indicate that 
the color -- color segregation is efficient and that the typical 
contamination amounts to $\sim$20-30\% (see also Pierini et al. 2003). 
\begin{acknowledgements}
We thank R. Gilli for useful discussion, J. Afonso for information
about the {\it Phoenix} survey and the referees for the constructive 
suggestions.
\end{acknowledgements}

\end{document}